\documentclass{sf2a-conf2024}
\usepackage{graphicx}
\usepackage{hyperref}
\usepackage[]{natbib}  
\usepackage{epstopdf}

\def\BibTeX{{\rm B\kern-.05em{\sc i\kern-.025em b}\kern-.08em
    T\kern-.1667em\lower.7ex\hbox{E}\kern-.125emX}}
\bibpunct{(}{)}{;}{a}{}{,}  

\begin{document}

\TitreGlobal{SF2A 2024}


\title{Impact of uniform rotation on the stochastic  \\ excitation of acoustic modes in solar-like oscillators}

\runningtitle{Short title here}

\author{L. Bessila}\address{Université Paris-Saclay, Université Paris Cité, CEA, CNRS, AIM, Gif-sur-Yvette, F-91191, France}
\author{A. Deckx van Ruys $^{1,}$}\address{Ecole polytechnique, Institut   Polytechnique de Paris, Palaiseau, France}
\author{V. Buriasco $^1$} 
\author{S. Mathis $^1$}
\author{L. Bugnet}\address{Institute of Science and Technology Austria (IST Austria), Am Campus 1, Klosterneuburg, Austria}
\author{R.A. García $^1$} 
\author{S. Mathur $^{4,}$} \address{Instituto de Astrofísica de Canarias (IAC), E-38205, La Laguna, Tenerife, Spain} \address{Universidad de La Laguna (ULL), Departamento de Astrof\'isica, E-38206 La Laguna, Tenerife, Spain}




\setcounter{page}{237}


\maketitle


\begin{abstract}
We evaluate the impact of the rotation on the stochastic excitation of acoustic (p) modes in solar-like pulsators. First, we derive the forced wave equation taking rotation into account and we compute the source terms, which inject energy into the oscillations. We make use of the Rotating Mixing Length Theory (R-MLT) to assess how the convective root mean square velocities are modified by the Coriolis acceleration. Finally, we use the stellar structure and evolution code MESA combined with the stellar pulsation code GYRE to show that the resulting modes amplitudes are inhibited by rotation. \end{abstract}



\section{Introduction}
Acoustic oscillations are stochastically excited by turbulent convection in solar-like pulsators \citep[e.g.][]{samadi_excitation_2001, belkacem_stochastic_2008}. Rotation and magnetic field seem to hinder modes detection \citep[e.g.][]{garcia_corot_2010, chaplin_predicting_2011}. \cite{mathur_revisiting_2019} witnessed that acoustic modes are not detected in more than 40 \% of solar-type stars from the \textit{Kepler} data. As these stars possess a convective envelope, which should drive the oscillations, we explore the hypothesis that the turbulent excitation source is too low to generate oscillations. Rotation and magnetic fields are known to modify convection \citep[e.g. ][]{chandrasekhar_hydrodynamic_1961}. In this work, we extend the theoretical model for the stochastic excitation of p modes \citep{samadi_excitation_2001} to include the effects of rotation, through its influence on convection.

\section{Stochastic excitation of acoustic modes with rotation}

\begin{figure}[ht!]
 \centering
 \includegraphics[width=0.6\textwidth,clip]{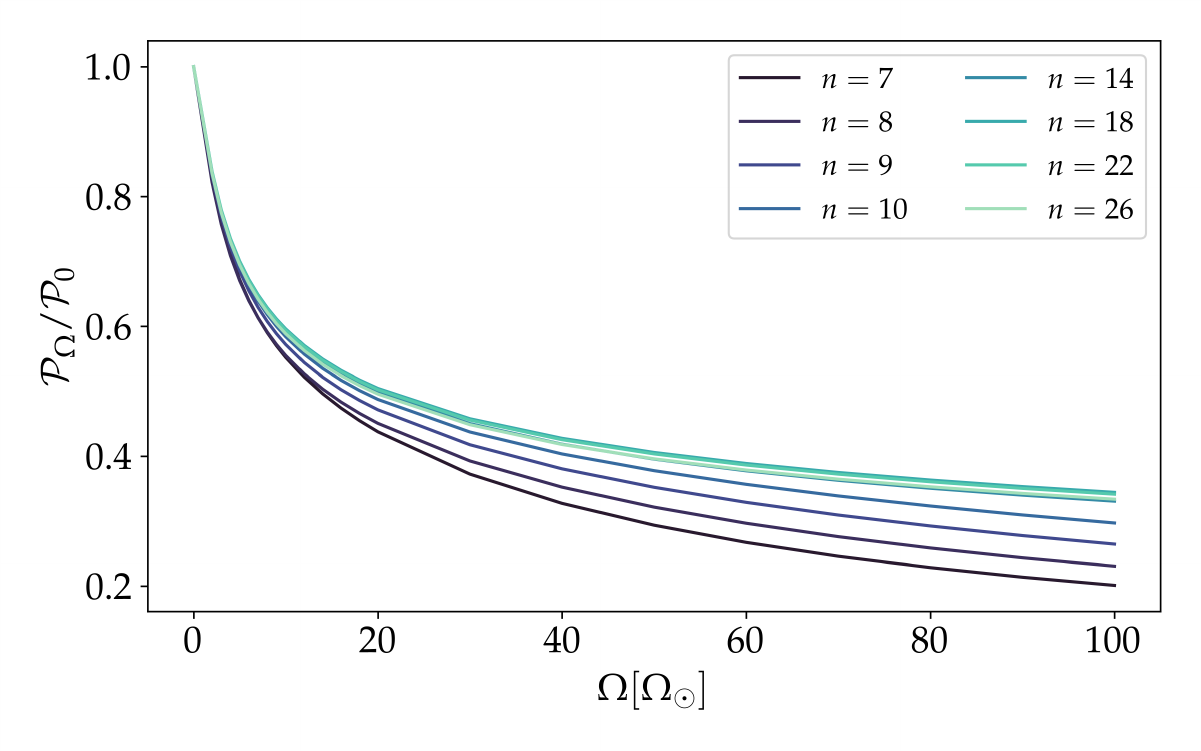}    
  \caption{We represent the power $\mathcal{P}_{\Omega}$ injected by turbulent convection into the acoustic modes as a function of rotation ($2 \pi/ \Omega_{\odot}=28$ days is the solar rotation rate) normalized by its value $\mathcal{P}_0$ in the non-rotating case. All the modes have the same $\ell=0$, and different values of radial order $n$.}
  \label{fig:rotation}
\end{figure}

\subsection{Stochastic excitation}
Even if various source terms can excite acoustic modes, \citep{samadi_excitation_2001, belkacem_mode_2009}, only the Reynolds stresses term is non-negligible in solar-like stars. Rotation indirectly modifies the stochastic excitation as it influences turbulent convection \citep[e.g.][]{brun_magnetism_2017}. As derived in \cite{samadi_excitation_2001}, the power injected into a given acoustic mode ($n,\ell, m$) is $$\mathcal{P}=\frac{16}{15 I^2} \pi^3 \int_{\mathcal{V}} d^3 x_0 \rho_0^2\left|\frac{\mathrm{d} \xi_{r,n,\ell}(r)} {\mathrm{~d} r}\right|^2 Y_{\ell,m} (\theta, \varphi) Y^{*}_{\ell,m}(\theta, \varphi)\hat{S}_R\left(r,\theta,\omega_0\right),$$ where $I$ is the mode inertia, $\rho_0$ the mean density, $\xi_{r,n, \ell}$ the radial component of the oscillation Lagrangian displacement, $Y_{\ell,m}$ the spherical harmonic of orders $\ell, m$. and $Y^{*}_{\ell,m}$ its complex conjugate. $\hat{S}_R$ is the contribution of the Reynolds stresses source: 
$\hat{S}_R\left(\omega_0\right)=\int \frac{d k}{k^2} E^2(k) \int d \omega \chi_k\left(\omega+\omega_0\right) \chi_k(\omega),$ where $E$ is the kinetic energy spectrum and $\chi_k$ is the eddy-time correlation spectrum. We find that a Gaussian time-correlation spectrum would not influence the excitation in a rotating framework, so we favour a Lorentzian function (Bessila et al. 2024, under review) since observations \citep{mathur_revisiting_2019} show a dependence of the excitation on rotation.

\subsection{Rotating convection}
To model rotating convection, we consider the single-mode approach of the Rotating Mixing-Length Theory, developed by \cite{stevenson_turbulent_1979} and \cite{augustson_model_2019}, which is based on the assumption that convection is dominated by the linear convective mode that carries the most heat \citep{malkus_heat_1954}. This model prescribes that the convective velocity scales like $\Omega^{-1/5}$. The R-MLT model succeeds in reproducing the results from local numerical simulations of rotating convection \citep[see e.g.][]{barker_theory_2014}.

\section{Numerical results}
 We assess the impact of uniform rotation on the stochastic excitation making use of the MESA stellar evolution code \citep{paxton_modules_2011, paxton_modules_2013, paxton_modules_2015, paxton_modules_2018, paxton_modules_2019, jermyn_modules_2023}, as well as the GYRE stellar pulsation code \citep[e.g.][]{townsend_gyre_2013}. We compute a solar-like model, with a mass $M = 1 M_{\odot}$ and a metallicity $Z=0.02$.
As shown in Fig. \ref{fig:rotation}, the power injected into the oscillations decreases when the rotation rate increases. For a $20 \Omega_{\odot}$ rotation period, the injected power diminishes by up to 50\%. In addition, all the modes are not equally influenced by rotation: low $n$ order mode amplitudes are more sensitive to this diminution.

\section{Conclusion}

We have included rotation in the theoretical model for the stochastic excitation of acoustic modes in rotating solar-like pulsators. We show that the power injected into the stellar oscillations can diminish by up to 50 \% for a $20 \Omega_{\odot}$ rotation rate, due to the impact of rotation on the convection. For a given $\ell$, low $n$ mode amplitudes are more inhibited.

\begin{acknowledgements}
L.B. and S.M. acknowledge support from the  European  Research Council  (ERC)  under the  Horizon  Europe programme  (Synergy  Grant agreement 101071505: 4D-STAR), from the CNES SOHO-GOLF and PLATO grants at CEA-DAp, and from PNPS (CNRS/INSU). While partially funded by the European Union, views and opinions expressed are however those of the authors only and do not necessarily reflect those of the European Union or the European Research Council. Neither the European Union nor the granting authority can be held responsible for them.\end{acknowledgements}

\bibliographystyle{aa}  
\bibliography{BESSILA_S01.bib} 

\end{document}